%% file: arxiv.tex
\documentclass[conference,compsoc]{IEEEtran}
\ifCLASSOPTIONcompsoc
  \usepackage[nocompress]{cite}
\else
  \usepackage{cite}
\fi
\ifCLASSINFOpdf
\else
\fi
\usepackage{algorithm}
\usepackage{algpseudocode}
\usepackage{amsmath}
\usepackage{amssymb}
\usepackage{enumitem}
\usepackage{pifont}
\usepackage{boxedminipage}
\usepackage{multirow}
\usepackage{tcolorbox} 
\usepackage{amsthm}
\usepackage{endnotes}
\usepackage{booktabs}
\usepackage{hyperref}
\usepackage{tikz}
\usetikzlibrary{shapes,arrows,positioning}
\newtheorem{definition}{Definition}

\newif\ifdraft
\drafttrue
\draftfalse

\ifdraft
\newcommand{\dimitra}[1]{\textcolor{blue}{\textbf{Dimitra:} #1}}
\newcommand{\note}[2]{\textcolor{blue}{\textsuperscript{\textbf{#1}} #2}}
\newcommand{\addition}[1]{\textcolor{blue}{#1}}
\newcommand{\kapil}[1]{\textcolor{purple}{\textbf{Kapil:} #1}}
\newcommand{\cedric}[1]{\textcolor{green!50!black}{\textbf{Cedric:} #1}}
\newcommand{\stavros}[1]{\textcolor{orange}{\textbf{Stavros:} #1}}
\else
\newcommand{\dimitra}[1]{\textcolor{blue}{}}
\newcommand{\note}[2]{\textcolor{blue}{}}
\newcommand{\addition}[1]{\textcolor{blue}{}}
\newcommand{\kapil}[1]{\textcolor{purple}{}}
\newcommand{\cedric}[1]{\textcolor{green!50!black}{}}
\newcommand{\stavros}[1]{\textcolor{orange}{}}
\fi

\newcommand{\registry}{\textsc{CoRe}}
\newcommand{\jfs}{\textsc{ZFS}}
\newcommand{\sjfs}{\textsc{ShieldZFS}}

\newcommand{\project}{\textsc{ShieldFS}}
\newcommand{\projectname}{\project}

\begin{document}
%\title{Bare Demo of IEEEtran.cls for\\ IEEE Computer Society Conferences}
\title{Securing Filesystems for Confidential Computing}

%\author{\IEEEauthorblockN{Michael Shell}
%\IEEEauthorblockA{School of Electrical and\\Computer Engineering\\
%Georgia Institute of Technology\\
%Atlanta, Georgia 30332--0250\\
%Email: http://www.michaelshell.org/contact.html}
%\and
%\IEEEauthorblockN{Homer Simpson}
%\IEEEauthorblockA{Twentieth Century Fox\\
%\and
%\IEEEauthorblockN{James Kirk\\ and Montgomery Scott}
%\IEEEauthorblockA{Starfleet Academy\\
%San Francisco, California 96678-2391\\


% compsoc conferences compared to traditional conferences), use this
\author{\IEEEauthorblockN{Dimitra Giantsidi\IEEEauthorrefmark{1},
Antoine Delignat-Lavaud\IEEEauthorrefmark{1},
C\'edric Fournet\IEEEauthorrefmark{1}, 
Jinnan Guo\IEEEauthorrefmark{2},
Heidi Howard\IEEEauthorrefmark{1},}
Tianjiao Huang\IEEEauthorrefmark{3},
Kapil Vaswani\IEEEauthorrefmark{4} and
Stavros Volos\IEEEauthorrefmark{1}
\IEEEauthorblockA{\IEEEauthorrefmark{1}Azure Research, Microsoft, Cambridge, UK}
\IEEEauthorblockA{\IEEEauthorrefmark{2}Imperial College London, UK}
\IEEEauthorblockA{\IEEEauthorrefmark{3}University of California, Irvine, USA}
\IEEEauthorblockA{\IEEEauthorrefmark{4}SPARC, Indian Institute of Science, India}}


\maketitle

\begin{abstract}
Confidential computing protects applications inside Trusted Execution Environments (TEEs), but it leaves storage vulnerable. Even with disk encryption, a malicious cloud provider can roll back, replay, fork, or tamper with disk state, breaking the integrity and freshness guarantees required by stateful applications. Existing solutions either assume trusted storage, incur high overheads, or push integrity logic into applications.

We present \projectname{}, a POSIX-compliant filesystem that provides end-to-end integrity and freshness for persistent storage in the confidential-computing threat model without requiring application changes. \projectname{} represents permissible filesystem states using succinct cryptographic commitments, maintained inside TEEs and replicated in a lightweight trusted registry. On-disk data structures, including a write-ahead log and a storage pool, are authenticated using hash chains and an embedded Merkle tree. \projectname{} utilizes transactions and copy-on-write to update persistent filesystem state and commitments atomically. The commitments are verified during reads, ensuring that rollback, replay, and equivocation attacks are detected even when the entire I/O stack is untrusted.

We implement the design by extending ZFS, yielding \sjfs{}. Evaluation with standard filesystem benchmarks and real-world workloads shows that \sjfs{} provides strong integrity and freshness guarantees with performance comparable to state-of-the-art filesystems.
\end{abstract}


% \fi
\IEEEpeerreviewmaketitle

\input{introduction}
\input{background}

\input{problem_motivation}
\input{overview}
\input{formal-model-definition}
\input{architecture}
\input{implementation}
\input{evaluation}

\input{related_work}

\section{Conclusion}
We presented \projectname{}, a POSIX-compliant filesystem that provides end-to-end integrity and freshness for persistent storage. \projectname{} turns filesystem structures that hold persistent state into authenticated data structures, and represents their permissible states using cryptographic commitments that are replicated in a lightweight trusted registry. Our analysis confirmed that \projectname{} fulfills the intended security guarantees under tampering, reordering, and replay attacks while maintaining practical performance levels. 

\vspace{0.1in}
\noindent \textbf{Software Artifact.} Our artifact is publicly available: 
{\url{https://github.com/dgiantsidi/ShieldFS}}.

\section*{Acknowledgments} This work would not be possible without support from  Eddy Ashton,  Amaury Chamayou,  Natacha Crooks,
Adrien Ghosn, Chris Jensen, Oleksii Oleksenko and the contributors of OpenZFS.

\ifCLASSOPTIONcompsoc
  %\section*{Acknowledgments}
\else
  %\section*{Acknowledgment}
\fi



%\bibliographystyle{IEEEtran}

{\footnotesize \bibliographystyle{IEEEtran}
\bibliography{sample_ieee}}

\end{document}

%% file: introduction.tex
\section{Introduction}
Cloud computing has become the default substrate for modern applications because it provides persistence, availability, and elasticity at scale. At the same time, regulatory and geopolitical pressures have elevated trust and data sovereignty to first-class concerns: organizations must ensure their data remains protected against external attackers and the cloud provider.

To address these concerns, technologies such as confidential computing have emerged. Confidential computing provides hardware-enforced isolation and remote attestation for computation and in-memory data processing using Trusted Execution Environments (TEEs), enabling applications to run securely even in the presence of compromised privileged attackers such as cloud administrators and service providers.

TEEs protect data during computation in memory, but not in stable storage (e.g., virtual or physical disks, or remote cloud storage). Even if confidential applications encrypt their state (using, e.g., dm-crypt or BitLocker with TEE-protected keys), an attacker that controls their storage system can still replay, roll back, delete, or fork data without being detected, and thus break their intended security. 

For example, a database running inside a TEE may read stale or tampered data from storage, leading to data corruption or loss of consistency. Similarly, TEEs relying on untrusted storage cannot ensure the freshness of their files, enabling an 
attacker to roll back critical changes.

Designing practical storage systems with strong integrity and freshness in this threat model is surprisingly hard.
Prior work has approached this problem by embedding integrity checks into application-level protocols~\cite{8418608}, but such designs sacrifice programmability, forcing developers to re-implement storage abstractions on a case-by-case basis,
instead of just trusting standard APIs such as POSIX.  
Lower-layer integrity-preserving block devices exist~\cite{chu2025rollbaccineherdimmunity}, but they incur substantial resource overheads or require specialized hardware and a redesign of the cloud storage backend~\cite{chrapek2025snvmeofsecureefficientdisaggregated}. Integrity-preserving filesystems have also been proposed~\cite{ipfs,8809505}, but either assume benign storage or fail to prevent rollbacks. In short, existing approaches either do not scale, or do not survive the attacker model relevant to modern clouds.

This leads to the question: 
{\it How can we build POSIX-compliant filesystems that ensure integrity and freshness under the confidential computing threat model while maintaining performance, scalability, and convenience?}

Our key principle is to organize the filesystem's persistent state as authenticated data structures~\cite{ads}, and to incrementally maintain their cryptographic commitments within protected TEE memory.
These commitments serve as succinct, verifiable representations of the filesystem current state, enabling the TEE to detect any deviation  
as it reads data from untrusted storage. To ensure commitments are persisted atomically, we leverage mechanisms such as transactions and copy-on-write deployed in several existing filesystems.
To ensure filesystems can survive TEE crashes and reboots, we register commitments at synchronization points with a trusted, fault-tolerant service also hosted in TEEs, and we consult this registry to bootstrap recovery from the filesystem's latest commitments. 

We implement this design by extending ZFS, a widely used open-source filesystem that already supports crash-consistency and integrity protection against benign data corruption. 
The resulting \sjfs{} runs inside TEEs---in our case SEV-SNP-protected VMs---and exposes a standard POSIX API to the applications running in that TEE. Our extensions involve minimal changes to ZFS on-disk layout and core logic, making them applicable 
in principle to similar filesystems. 

We also implement a commitment registry by adapting CCF~\cite{10.14778/3626292.3626304}, 
a framework for securely replicating a key-value store
across TEEs, so that it can serve a large number of filesystems with minimal resource overheads.

We optimize it to achieve low-latency updates (0.68 ms), which is critical for synchronous workloads. 
Prior rollback-protection mechanisms incur higher latencies: 2 ms for ROTE~\cite{203712}, 2.5 ms for Nimble~\cite{angel2023nimble}, and $\approx$100 ms when using SGX hardware counters~\cite{203712, 227798}.

We evaluate \sjfs{} along two dimensions. First, we perform a security analysis based on precise models for the filesystem and the attacker, showing that it detects tampering, rollback, replay, and equivocation attacks even when the cloud provider is fully malicious. Second, we conduct a performance evaluation using filesystem and real-world workloads, showing that, compared to ZFS, \sjfs{} achieves low overheads ($<$10\%) for most workloads.
The commitment registry can be shared by as many as 25 write-heavy filesystems while sustaining low registration latency.

In summary, we make the following contributions:

\begin{itemize}
\item We present \projectname{}, a generic filesystem architecture to ensure integrity and freshness 
on top of untrusted storage backends, while remaining agnostic to (most of) the underlying filesystem implementation.
\item We specify durability, crash-consistency, integrity, and freshness for POSIX filesystems. 

\item We design and implement \sjfs{}, a POSIX-compliant filesystem that builds on ZFS and enforces end-to-end integrity.

\item We design and implement \registry{}, a secure, lightweight, low-latency registration service that builds on CCF. 

\item We provide a security analysis of \sjfs{} and \registry{} in the confidential computing attacker model. 

\item We evaluate \sjfs{} and \registry{} as drop-in replacements for existing production filesystems, 
showing that they achieve integrity and freshness with low overheads and full compatibility with existing applications.
\end{itemize}

%% file: background.tex
\section{Background}
\noindent{\bf Confidential Computing.}
Confidential computing enables the execution of code and data within trusted execution environments (TEEs) that protect them against unauthorized access or modification, even from privileged software such as operating systems or hypervisors. 

In addition to isolation, confidential computing also supports remote attestation, which allows a remote party to verify the integrity of a TEE before provisioning it with sensitive data or code. TEEs can ask the hardware to sign a message together with a digest of its initial configuration (code and data) with a key only accessible to the hardware. The signature is backed by a platform certificate issued by the hardware provider. Through remote attestation, a client can verify a TEE’s trusted computing base (TCB) and hardware platform.

Modern CPUs from Intel, AMD, and Arm provide confidential computing capabilities. AMD Secure Encrypted Virtualization (SEV-SNP)~\cite{amd_sev_article}, Intel Trusted Domain Extensions~\cite{intel_tdx}, and Arm Confidential Compute Architecture~\cite{arm_cca} provide VM-based TEEs that isolate full-fledged virtual machines. Accelerators such as NVIDIA GPUs~\cite{nvidia_cc} also support confidential computing capabilities.

\noindent{\bf POSIX Filesystems.}
The POSIX filesystem API~\cite{posix_standard} is a standard interface for applications to interact with persistent storage through operations such as \texttt{open}, \texttt{read}, \texttt{write}, \texttt{fsync}, \texttt{rename}, and \texttt{unlink}. POSIX specifies the effects of these operations
and their durability. In particular, it requires that reads account for preceding writes within a process, and that synchronization operations such as \texttt{fsync}, \texttt{sync}, and \texttt{syncfs} ensure that preceding writes have been persisted to stable storage.

However, POSIX does not specify how filesystems should implement crash recovery, nor 
how they should behave when data is corrupted. It assumes a benign storage system in which data is persisted correctly or fails in detectable ways. 

\noindent{\bf Crash-Consistent Filesystems.}
Most filesystems implement some form of crash-consistency using logging or journaling techniques. Updates are first applied to in-memory data structures and then persisted to disk. To ensure that the filesystem can recover after a crash, updates are typically recorded in a write-ahead log (WAL) or journal before being applied to their final locations. During recovery, the filesystem replays the log to reconstruct a consistent in-memory state.

Many filesystems, including ZFS~\cite{openzfs_github} and btrfs~\cite{btrfs-trees-doc}, additionally employ copy-on-write (CoW) semantics. Instead of modifying blocks in place, updates create new versions of blocks at fresh locations. Metadata is updated to point to the new versions, and older versions remain intact until they are garbage-collected. This approach enables atomic updates and simplifies recovery, since the filesystem can switch between consistent versions by updating a small set of root pointers.

\noindent{\bf Integrity Protection.}
Modern storage systems employ various integrity mechanisms to protect against data corruption caused by hardware faults or software bugs. At the block level, data is often protected using checksums or cyclic redundancy checks (CRCs), which are stored alongside the data and verified on reads. Some filesystems, such as ZFS, extend this idea by organizing blocks into Merkle-tree-like structures, where each block includes a checksum of its children. This allows early detection of a large class of corruptions. At the block-device layer, dm-integrity~\cite{dm_integrity} in Linux provides cryptographic integrity for storage devices by maintaining 
an authentication tag for each block. 

These mechanisms protect against benign faults such as bit rot, misdirected writes, or transient hardware errors, but they are not designed to protect against active adversaries that control the storage layer. 
Dm-verity~\cite{dm-verity} provides integrity verification for whole immutable disk images using Merkle trees, but it does not support updates (which would incur substantial write amplification).

%% file: problem_motivation.tex
\section{Threat Model}
\label{sec:threat-model} 

We adopt the usual threat model for confidential computing~\cite{10.1145/2799647, 277820, 305248, 8418608}: 
the Trusted Computing Base (TCB), including all code and data in memory 
for the operating system and application running inside the TEE, cannot be observed or tampered with. 
Hence, side-channel attacks and DRAM-based attacks (e.g., Rowhammer, or physical memory interposer) 
and their mitigations are out of scope.

Code running inside TEEs is trusted to correctly implement \projectname{} 
and its applications, as well as the registry service. As usual, their TCBs can be validated 
through remote attestation, and independently reviewed and audited.
This excludes, for example, supply chain attacks where an attacker injects malicious code into the TEE software stack, or roll it back to some vulnerable code version. 

Conversely, we assume a powerful adversary that controls and tampers with the entire software system stack in the cloud, including operating systems, drivers, and hypervisors, as well as all hardware components excluding the TEEs~\cite{10.1145/3626718, 9833660, 227798, 278332}. 
This represents attackers capable of executing storage and network attacks, 
meaning they can access, corrupt, swap, drop, record, inject, or replay any data across their interfaces.
Rollback and forking attacks are possible by crashing TEEs and restarting them from stale snapshots of their persisted state, or executing multiple TEEs.

In this threat model, commodity filesystems running 
within TEEs are vulnerable to a broad range of attacks. We distinguish between {\it online} and {\it offline} attacks. Online attacks are performed while the system is up and running. In practice, such attacks are easier to detect. For example, the filesystem can store integrity metadata in protected memory and validate data fetched from disk against this metadata. Offline attacks are performed while the TEE is not running. Such attacks are harder to detect, since the adversary can rewrite the TEE's persistent state before the next execution.
Next, we illustrate offline attacks through concrete examples.

\noindent\textbf{(1) Rollback attacks.} Consider a block device with dm-integrity enabled. An adversary controlling the storage backend can record a valid disk state, and later replace the whole disk with this snapshot. Since both data and authentication tags are consistent, the rollback goes undetected. Applications may observe stale but well-formed data, breaking durability guarantees of operations such as \texttt{fsync}.

\noindent\textbf{(2) WAL rewriting.} An adversary can tamper with the WAL between executions by replaying older log entries, reordering existing ones, or selectively dropping records. 
As long as the tampering preserves internal consistency 
to avoid detection, recovery will succeed, and the filesystem will reconstruct a state that reflects an arbitrary prefix or permutation of previously committed operations, allowing an attacker to silently erase, revert, or reorder durable updates.

\noindent\textbf{(3) Forking attacks.} An adversary can present different views of the persistent state to different instances of the TEE, and direct different 
clients to instances that will independently apply updates
to their forked states, leading to inconsistencies and violations of application invariants.

%% file: overview.tex
\section{\projectname{} Overview}\label{sec:overview}
\noindent{\bf Filesystem Assumptions.} Our design builds on any filesystem with built-in mechanisms against benign data corruption. 
Specifically, we assume that the underlying filesystem provides:
\textbf{(1) Rooted persistent state.} The filesystem organizes persistent data and metadata as acyclic data structures rooted at a small set of blocks.
\textbf{(2) Copy-on-write updates.} The filesystem uses copy-on-write semantics to update persistent data and metadata, ensuring that any modification results in a new version of the data structure.
\textbf{(3) Deterministic recovery from WAL.} The filesystem uses a write-ahead log (WAL) to record all updates (data and metadata) before they are applied, and recovery proceeds by replaying committed WAL entries to restore a consistent state after a crash.
Such mechanisms are common in modern filesystems, e.g., ZFS~\cite{openzfs_github}, Btrfs~\cite{btrfs-trees-doc}, and ReFS~\cite{refs}.

\noindent{\bf Key idea.} The key idea behind \projectname{} is to turn both the filesystem stable storage and WAL into authenticated data structures, and to represent the permissible persistent states of these data structures using \textit{cryptographic commitments}. 
A commitment is a succinct record that collects cryptographic hashes or MACs 
to authenticate a particular state of the whole filesystem.  
In our case, a commitment consists of authenticators that represent the state of the WAL and the root of stable storage. 
The WAL head and tail authenticators capture all updates that have been durably recorded but not yet propagated to the main storage, while the storage root authenticator summarizes the fully persisted filesystem state. 
In combination, they uniquely authenticate the complete logical state of the filesystem at a given point in time, and thus suffice to verify its integrity and freshness.

\begin{figure}[t]
    \centering
    \includegraphics[width=\columnwidth]{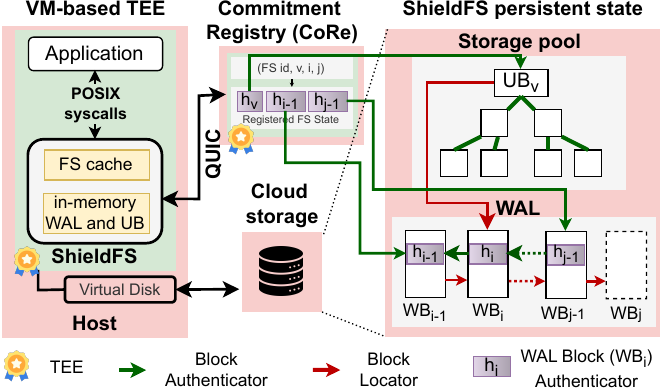}
    \caption{System overview}
    \label{fig:system_overview}
\end{figure}

\noindent{\bf System overview.} Figure~\ref{fig:system_overview} illustrates the high-level architecture of our system. Green boxes denote trusted components, such as TEEs and authenticated network links; all other elements are considered untrusted.

Our system comprises three main components. A confidential VM runs the application within a TEE and mounts \projectname{} 
to manage its persistent state, with the programmability, consistency, and durability of any POSIX filesystem.
Applications interact with \projectname{} through standard 
system calls (e.g., \texttt{read}, \texttt{write}, \texttt{fsync}), as modeled in \S\ref{sec:file_system_model}.
\projectname{} is configured to persist data in cloud storage (local or remote) using standard lower-level block I/O interfaces; this storage is untrusted from a security viewpoint, but still critical for availability. 
\projectname{} also communicates with a \textit{commitments registry} over a mutually attested secure channel to store and retrieve its commitments.

\projectname{} verifies all reads from persistent storage using cryptographic authenticators stored alongside the data, and updates these authenticators on writes.
When an application performs a \emph{synchronous update}, for example by writing data and invoking \texttt{fsync} to request 
durability, \projectname{} persists the update in the WAL, then computes and registers a new commitment 
for the updated WAL. Once the registry acknowledges the update, \projectname{} completes the \texttt{fsync} and returns control to the application. 
In the background, \projectname{} also persists asynchronous updates and flushes the WAL to stable storage, which also involves registering new commitments. 

Upon recovery or remount, \projectname{} authenticates to the registry service (using its TEE attestation), and then fetches the registered commitments representing its last known good state. 
It then verifies that the persistent state on disk matches the expected state authenticated by the commitment. If verification succeeds, \projectname{} recovers to the corresponding state; otherwise, it detects an integrity or freshness violation and aborts recovery, preventing the application from observing corrupted or stale state. 
(The application owner might then decide to recover from a backup, similarly authenticated by a commitment, and accept the risk of losing recent updates.)

Compared with secure device mappers, which operate at the block layer and must persist authenticators for every block write, our design limits the need for trusted persistent state to the succinct commitments held by the registry---a few hundred bytes per filesystem, and requires significantly fewer synchronizations.
All other intermediate authenticators coexist with data in untrusted storage.
As shown in our evaluation (\S\ref{sec:evaluation}), this yields strong integrity and freshness guarantees with low performance overheads.

%% file: formal-model-definition.tex
\section{Modeling Integrity and Freshness}
\label{sec:file_system_model}

We model a filesystem by its interaction
with an application. The application program consists 
of multiple threads that access the filesystem 
using a POSIX API~\cite{posix_standard}. We keep the program implicit, and model its system calls 
and their results as {\em events} that reflect what we would experimentally observe 
in a system-call trace. All events are atomic, but some operations may be split into multiple events, e.g., reading a large file may yield a series of read events, each returning a chunk of the file.

\begin{definition}[Events]
Events are atomic interactions 
between an application and the filesystem, ranging over
\begin{itemize} [left=0pt, labelsep=0.5em, noitemsep, topsep=0pt]
    \item \emph{Reads} $r$ that return information about the filesystem state;
    \item \emph{Updates} $u$ (write, extend, rename) that modify this state;
    \item \emph{Synchronization events} $s$ (fsync, sync, syncfs) that commit updates from this state to stable storage.
\end{itemize}
\end{definition} 

\noindent For example, the event $\mathit{read}(inode, addr, data)$ states that reading at a given address returned some data; 
it may instead return an error, e.g., when reading after the end of file. Similarly, $\mathit{write}(inode, addr, data)$ updates 
the state of an open file in memory; and $\mathit{fsync}(inode)$ confirms that all pending updates on a given inode have been persisted to stable storage. 

A \emph{session} captures all events between mounting and unmounting (or crashing) the filesystem, and the set of update events that have persisted. 

\begin{definition}[Session]\label{def:session}
A session $S = (\vec t,\vec u)$ consists of 
a \emph{live sequence} ($\vec t$) of all events in the session  
and a \emph{commit sequence} ($\vec u$) that
collects, for every inode, a prefix of the updates for this inode in $\vec t$,
including \emph{at least} every update followed by a synchronization for this inode in $\vec t$.
\end{definition}
Hence, reflecting POSIX semantics, the commit sequence $\vec u$
coincides with the live sequence $\vec t$ at least up to the last synchronization event 
of every inode, but it may include more recent updates.
On the other hand, the interleavings of updates between different inodes may differ.

The commit sequence $\vec u$ indicates the updates that have been persisted  
before the system crashes (including the WAL). Cleanly unmounting the filesystem is the special case where all updates in $\vec t$ have been persisted.
On the other hand, to model recovery, 
the live and commit sequences of a session may start 
with the commit sequence recovered from a prior session.

Next, we relate reads to the filesystem state 
after applying a sequence of updates.

\begin{definition} The filesystem state $\sigma$ is defined by 
\begin{itemize} [left=0pt, labelsep=0.5em, noitemsep, topsep=0pt]
\item $\emptyset$: its initial empty state (with no files or directories); 

\item $\mathit{Update}(\sigma, u)$, a function that computes the new state after executing the update event $u$; and 

\item $\mathit{Read}(\sigma,r)$: a predicate that states the read event $r$ reflects the current state.
\end{itemize}
\end{definition}

\noindent  For example, $\sigma$ may be a map from inodes to lists of blocks, and 
$Read(\sigma,read(1, 3, \sigma[1,3]))$ 
may state that a read returns one of these blocks to the application. We write $\mathit{Update}(\sigma, \vec u)$ for the state after applying a sequence of updates~$\vec u$. 
We say that a read event $r$ follows from these updates when $\mathit{Read}(\mathit{Update}(\emptyset,\vec u), r)$.

We can reason about properties that hold for all possible sessions
that may be observed when interacting with the filesystem. For example, the definition below captures our intuition 
that POSIX serializes all live events: 

\begin{definition}\label{def:posix_sync}
  A session is POSIX-compliant when every read event in $\vec t$ follows 
  from the preceding update events in $\vec t$.  
\end{definition}

The details of individual system calls are out of scope for this work.
SibylFS~\cite{RidgeSibylFS2015} defines a formal model of POSIX compliance
that covers all allowed behaviors arising from any sequence of 21 system calls in a session.
However, persistence across crashes is not covered.

While mainstream filesystems ensure each session is POSIX-compliant (Definition~\ref{def:posix_sync}), they differ in their crash-consistency semantics, i.e., in the commit sequences that may be used for recovery (Definition~\ref{def:session}).
The POSIX standard underspecifies the contents and ordering of updates in persistent storage and, for example, {\it ext4} supports a weaker form of consistency~\cite{10.1145/2872362.2872406} whereas 
log-structured and journaled filesystems often provide stronger guarantees (allowing fewer potential commit sequences).

In this work, our focus is not on crash-consistency per se 
(that is, what the commit sequences may be for a given filesystem), 
but rather on its secure implementation in the confidential computing threat model
(that is, ensuring the integrity and freshness of the commit sequence used for recovery).

In the absence of corruption, POSIX implies strong integrity and freshness guarantees. For example, 
an application that completes an fsync after writing to a given file is guaranteed that, if the filesystem later crashes and recovers, the recovered file contents will include its update. To reflect events that can corrupt filesystem state, we introduce \emph{havocs} as a special kind of update events recorded in sessions.

\begin{definition} \emph{Havoc events} range over 
\begin{itemize} [left=0pt, labelsep=0.5em, noitemsep, topsep=0pt]
    \item Data corruptions 
    \emph{Corrupt(inode, addr)}, causing read events to return errors, % ($\bot$), 
    e.g., after failing to verify a checksum. 
    \item Data forgeries 
    \emph{Forge(inode, addr, data)}, causing read events to return this updated data, e.g. after corrupting the data 
    and adjusting its checksum, or replaying stale data.
\end{itemize}
\end{definition}

Crashes are not considered havocs, since our model already captures crashes at any point in a session, 
followed by a recovery from a commit sequence that may cause some uncommitted updates to be discarded. 
In the confidential computing threat model, attackers can introduce arbitrary havocs at any time,
including between a crash and a subsequent recovery, modeled as havoc events at the start of the recovery session. 

\begin{definition}[Integrity]
A filesystem provides integrity when
every read follows from preceding updates in $\vec t$, 
including initial updates from the commit sequence of a prior session
and data corruptions, but excluding data-forgeries. 
\end{definition}
 
\begin{definition}[Freshness]
A filesystem provides freshness when 
every read follows from all preceding updates in $\vec t$, 
including initial updates from the commit sequence of the last session
and data corruptions, but excluding data-forgeries.
\end{definition}

This second property is stronger and more precise, as it directly matches Definition~\ref{def:session} within every session, 
up to the fail-safe detection of any data corruption or forgeries, 
and also guarantees
the correct chaining of every committed update across sessions. It is our target for 
\projectname{}.

%% file: architecture.tex
\newcommand{\wb}{\mathit{WB}}
\newcommand{\ub}{\mathit{UB}}
\section{Shielding ZFS}\label{sec:architecture}
We now describe the architecture of \projectname{} for achieving integrity and freshness (\S\ref{sec:file_system_model}) in our threat model (\S\ref{sec:threat-model}).
We first review selected features of ZFS that \projectname{} depends on, and then describe our extensions to realize the ShieldFS design, yielding \sjfs{}. Although ZFS is a convenient baseline, ShieldFS also applies to other modern transactional filesystems such as btrfs and ReFS, which similarly combine write-ahead logging, copy-on-write updates, and recovery from a small set of root blocks.

\label{sec:shielded-jfs}
\subsection{Volatile State vs Persistent State} 
In \jfs{}, the filesystem state consists of blocks in stable storage together with volatile state in memory.
At the behest of the application, reads are served first from an in-memory cache~\cite{gregg2012zfs-arc}, if possible, and otherwise by reading blocks from disk. Similarly, updates are first applied in-memory, then eventually persisted by writing dirty blocks to disk, either as a result of cache eviction or to complete synchronization requests. Hence, volatile state is usually ahead of persistent state. 

From a security viewpoint, volatile state and its operations are hardware-protected by the TEE, so we can ignore their details (e.g., 
their in-memory synchronization mechanisms, and their adaptive cache data structures and policies) and focus on authentication as data is read back from persistent storage: by induction, as long as all these reads are correct,
ZFS always holds in memory the latest, correct state of the filesystem 
and will issue correct writes to persistent storage.

To reduce the latency of synchronous writes, and to increase locality, \jfs{} 
first persists all updates in a Write-Ahead-Log (WAL)\footnote{In \jfs{}, the WAL is referred to as the ZFS Intent Log (ZIL)} before eventually applying them to the main storage pool, and it reads back the WAL only for recovery after a crash, in order to reconstruct a volatile state that accounts for all committed updates. As detailed below, \sjfs{} separately authenticates these two forms of storage. 

\subsection{Storage Pool}

\emph{Blocks} are the basic logical unit of storage. 
\jfs{} employs different types of blocks for storing data, metadata, WAL updates, etc. 
In addition to its payload, each block has a header that includes its type, identifier, and size.
For our purpose, we need to understand how \jfs{} organizes and locates 
blocks in persistent storage, i.e., what constitutes a \emph{block pointer}, how they are persisted,
and how they are dereferenced. 

Each block is persisted in a \emph{storage pool} assigned to the filesystem. 
Accordingly, a \emph{block locator} provides relevant information to fetch a block from the pool, including the physical page locations on disk, or a virtual device identifier and an offset, or sub-locations in case the block is fragmented and/or replicated. Locators also include information for resource management, such as reference counting. 
In addition to locators, block pointers feature metadata about the target, notably a version number and checksum. 

The storage pool forms a graph, with blocks as nodes and block pointers as edges. Some of these nodes serve as entry points (or roots), each representing a different filesystem version, snapshot, or clone in stable storage.  
In particular, to mount a filesystem, \jfs{} first fetches a special type of block that summarizes its whole state, called an \emph{uberblock}, then recursively dereferences its block pointers to resume from this state in memory. 
For consistency, \jfs{} supports atomic updates using \emph{copy-on-write} semantics: the new version of a block
is first written to a free location in the pool. This returns an updated in-memory block pointer with a new locator.
This pointer can in turn be persisted by transitively updating parent blocks, so that they point to the updated version,
without invalidating existing pointers to prior versions.
Hence, updating the only pointer to a sub-graph atomically switches between consistent versions of this sub-graph.  

In particular, \jfs{} regularly persists new versions of the uberblock (keeping a few recent versions at fixed locations), each recursively pointing to the blocks that represent the latest consistent state of the whole storage pool. 
Assuming (for now) that uberblock updates are themselves atomic, this ensures that recovery can proceed from the consistent state reachable from the uberblock with the latest version number. 

\begin{figure}[t]
\centering
\small
\begin{tabular}{@{}l|l@{}}
\hline
$B_u$ &  Block in storage pool \\
$\ub_{i,v}$ & Uberblock version $v$ with head index $i$ \\
$\wb_k$ &  WAL block $k = 0, 1, \dots, i, \dots, j$ \\
$i \leq j$ &  WAL block head and tail indexes \\
$\ell_{u}, \ell_{k}, \ell_{v} $ & Physical locators \\ 
$H(\cdot)$ & Collision-resistant hash function \\ 
$h_{u} = H(B_u)$ & Authenticator of a block in storage pool  \\    
$h_{i,v} = H(\ub_{i, v})$ & Authenticator of an uberblock version \\
$h_{k} = H(h_{k-1}||\wb_k)$ & Authenticator of the WAL $(\wb_u)_{u\leq k}$ \\
$\mathit{FS}_{id}$ & Unique identifier of a filesystem   \\
$\mathit{head}_{i,v} = {}$ & Commitment to the storage pool \\ 
$\quad i,v,h_{i,v},h_{i-1}$ & \quad after applying all updates $(\wb_u)_{u<i}$  \\
$\mathit{tail}_j = j, h_{j-1} $ & Commitment to all updates $(\wb_u)_{u<j}$ \\
$\mathit{FS}_{id} \mapsto \mathit{policy}, $ &
Registered state of a filesystem \\ 
$\quad  \mathit{head}_{i_0,v}, \mathit{head}_{i_1,v+1}, $ \\ $\quad \mathit{tail}_j$ \\
\hline
\end{tabular}
\caption{Main notations in \S\ref{sec:architecture}}\label{fig:notations}
\vspace{-3ex}
\end{figure}

\textbf{Shielding the Storage Pool} requires authenticating every block read. To this end, ZFS already embeds in every block pointer an \emph{authenticator} that stores a checksum of the block it points to. 
\sjfs{} replaces this checksum with a cryptographic hash, such as SHA-256, that serves as a strong authenticator of the block content. Using a collision-resistant hash function ensures that any contents read will match the contents written at the time the block pointer was updated. And using a cycle-resistant hash function (another common cryptographic security assumption) ensures that the storage pool forms an acyclic graph---intuitively, the authenticator of a block that includes pointers to other blocks is computed by hashing their previously-computed authenticators as part of its contents, and the resulting hash cannot be predicted in advance (at least in the random-oracle model). 

To verify the integrity of the storage pool, \sjfs{} performs the following checks:
{\bf (1) When loading a block given a block pointer,} compute the hash of its retrieved contents, and compare it with the authenticator stored in the pointer.
If they match, add the block to the cache (or any other in-memory structure); otherwise, discard the block and return a `corrupted data' error. As before in ZFS, the authenticator can also be used for error correction when retrieving a block replicated at multiple locations, as long as at least one copy is intact. 
{\bf (2) When evicting a dirty block,} compute the hash of its stored contents 
and, once the write completes, return an updated block pointer that includes both its new locator and its new authenticator.
This pointer can then be used (as before) to update parent blocks.

Authenticators ensure basic atomicity of block updates, inasmuch as a partial write will cause an error when reading back the block, rather than return a partially-updated block. 
In particular, {\bf (3) when saving a new version $v$ of the uberblock,} written $\ub_{i,v}$, any dirty blocks it points to must have been already persisted, and its updated 
authenticators form a commitment to the whole storage pool state.
\sjfs{} computes an uberblock authenticator $h_{i,v} = H(\ub_{i,v})$ that succinctly represents this commitment, and registers it as the next stable state of the filesystem (see \S\ref{subsec:protocols}). 
Once registration completes, \sjfs{} writes $\ub_{i,v}$ to the next available uberblock location on disk using \jfs's two-phase update. 

\subsection{Write-Ahead Log}

The WAL (also known as the ZFS Intent Log or ZIL) is an append-only crash-consistency log~\cite{kernel-ext4-journal,zfs-ondisk-format,btrfs-trees-doc}.
It aims to minimize write amplification when the application requires persistence of recent updates.
Instead of waiting for the filesystem to write and possibly re-organize many locations in the storage pool (including metadata updates, disk-page allocations, etc), these updates are written in bulk into WAL blocks, 
so that they can be replayed after a crash. Asynchronously, \jfs{} integrates these updates to the main storage pool and recycles their WAL blocks.   

The WAL logs the full sequence of filesystem updates, grouped in successive blocks $(\wb_k)_{k \geq 0}$. By convention,
we denote by $i$ the block sequence number at the head of the WAL (that is, the first block to be replayed)
and by $j$ the block sequence number at its tail (that is, the next block to be appended to the WAL).
The WAL logs sufficient information to replay updates during recovery: 
a type (e.g. create, write, truncate), 
the identifier of the updated object,
the range of data involved (offset and length) and 
the data being written, if any. WAL block headers also include the pre-allocated \emph{locator} of the next block, 
enabling its traversal given the locator $\ell_i$ of its head block, 
as well as a checksum of the block contents.  

\medskip\noindent{\bf Shielding the WAL.} To ensure integrity and freshness of the WAL at recovery time,
we equip it with authenticators, 
essentially building a hash chain over its blocks: 
\begin{equation}\label{equation:authenticator}
h_0 = H(0 || \wb_0) \qquad  h_{k} = H(h_{k-1} || \wb_{k}) \mbox { for } k > 0
\end{equation}
Hence, $h_k$ is a cryptographic commitment to the whole sequence of updates recorded in $\wb_0, \wb_1, \ldots, \wb_k$.

\jfs{} already keeps track of the block indexes $i$ and $j$ at the head 
and tail of the WAL, and of all intermediate locators $\ell_{i}, \ell_{i+1}, \dots, \ell_{j-1}$.
\sjfs{} additionally keeps in memory all intermediate authenticators $h_{i-1}, \dots, h_{j-1}$ so that
we can both verify the integrity of the WAL starting from its head
and extend the hash chain with new blocks appended at its tail.
Operations on the WAL are refined as follows. 

{\bf (1) Appending at the tail.} To fulfill synchronization requests, 
\jfs{} persists updates as the contents of a new in-memory WAL block $\wb_j$.
It pre-allocates a physical location on disk for the next block, which yields the next locator  $\ell_{j+1}$.
It writes this locator in the header field reserved for it in $\wb_j$.
In addition, \sjfs{} computes the authenticator $h_j$ from $h_{j-1}$ and the contents of $\wb_j$ using~\eqref{equation:authenticator}. 
It persists $\wb_j$ at its pre-allocated location $\ell_j$.
It registers the new tail authenticator $\mathit{tail}_{j+1} = (j+1, h_{j})$ (see \S\ref{subsec:protocols}).
Finally, it updates its in-memory state with the new tail index $j+1$, authenticator $h_j$, and locator $\ell_{j+1}$, 
and it signals completion of all synchronization requests pending on the updates in $\wb_j$.

{\bf (2) Propagating updates to the main pool.}
This is done incrementally and asynchronously using in-memory representations of the updates 
committed to the WAL, without any change to \jfs{}. 
Once every update in the head block $\wb_i$ has been persisted in the storage pool
by writing back new copies of the corresponding blocks, 
we increment the head index. 

Later, when writing a new version of the uberblock $\ub_{i,v}$ that commits all these updates to stable storage, 
we will use the latest in-memory head block index $i$ to include the WAL head locator $\ell_{i}$ in this uberblock, and to register it together with the prior WAL head authenticator $h_{i-1}$ (see \S\ref{subsec:protocols}). 
Finally, blocks in the WAL that are no longer reachable from the locators of any uberblock can be marked for garbage collection. 

\subsection{Recovery}\label{subsubsec:recovery}

We explain how \sjfs{} mounts a filesystem, possibly following a crash. 
Anticipating the protocols in \S\ref{subsec:protocols}, we assume we are given a commitment 
$(\mathit{head}_{i,v}, \mathit{tail}_j)$ that represents the current state of the filesystem, and we use it to 
bootstrap the verification of the storage pool and the WAL.

We write $\ub_{i, v}$ for version $v$ of the uberblock, 
which includes the locator $\ell_{i}$ of the head of the WAL, 
and $h_{i,v} = H(\ub_{i,v})$ for its authenticator. 
We use $\mathit{head}_{i,v} = i,v,h_{i,v},h_{i-1}$
to record the logical head of the WAL, including the indexes $i$ and $v$, $h_{\ub_{i,v}}$, 
and the authenticator of the preceding block $\wb_{i-1}$ in the WAL 
(used to continue the hash chain computation given its head block $\wb_i$).
We use $\mathit{tail}_j = j, h_{j-1}$ 
to record the tail of the WAL, including the authenticator of the last block appended to the WAL $\wb_{j-1}$.
In combination, $\mathit{head}_{i,v}$ and $\mathit{tail}_j$ record the full persisted state of \sjfs{}.
(See Figure~\ref{fig:notations} for a summary of notations.)

Recovery starts by reading candidate uberblocks from fixed locations on disk, which yields 
(at most) the two uberblocks $\ub_{i_0,v}$ and $\ub_{i_1,v+1}$ with the highest persisted versions $v$ and $v+1$, respectively.
With plain \jfs{}, recovery would either use $\ub_{i_1,v+1}$ after checking it is well-formed, or use $\ub_{i_0,v}$, and one of the two would succeed in the absence of data corruption.
With \sjfs{}, we first recompute their uberblock authenticators $h_{i_0,v}$ and $h_{i_1,v+1}$ by hashing their contents,
and we select the one that matches the trusted authenticator in $head_{i,v}$ 
(otherwise returning a `corrupted uberblock' error).
We can now use the authenticated uberblock to reload any other block in the storage pool, following block pointers 
and verifying their authenticators, as explained above. 

Next, we need to authenticate and replay the WAL.  
This step is skipped if the WAL is empty, i.e., $i \geq j$, which is the case when remounting after a clean unmount.
If $i < j$, starting from the WAL head block locator $\ell_i$ in the uberblock
and from index $k = i$ and authenticator $h_{i-1}$ in $\mathit{head}_{i,v}$, 
we iterate on the WAL blocks: 
while $k < j$,
we read back $\wb_k$, recompute $h_k$ from $h_{k-1}$ and $\wb_k$, 
and increment $k$. Once we have reached the tail, we verify that the computed authenticator
$h_{k-1}$
matches $h_{j-1}$ recorded in $\mathit{tail}_j$, and otherwise we report a `corrupted WAL' error.

Once authenticated, the WAL is replayed in memory (as in \jfs{}) which restores the volatile state ahead of the recovered storage pool authenticated by the trusted commitment, and we are ready to resume normal filesystem operations.

\subsection{Extensions to other filesystems}

Our \project{} approach can be applied with minimal changes to other filesystems. 
btrfs uses a copy-on-write update model with a root pointer (the superblock) and a WAL (the log tree). We can modify extent pointers to include authenticators (which already support CRC32 based checksums), and we would register the authenticator of the superblock along with the head and tail authenticators of the log tree.
Similarly, ReFS uses a copy-on-write B+ tree structure with a root pointer (the superblock) and a WAL (the log file). We can extend ReFS's block pointers to include authenticators, and register the authenticator of the superblock along with the head and tail authenticators of the log file.
Extensions to distributed filesystems such as CephFS would require additional mechanisms to ensure integrity across multiple nodes, but the core principles of authenticating block pointers and registering root and log authenticators remain applicable.

\subsection{Commitments Registry (\registry{})} \label{sec:registry} \label{subsec:protocols}

The registry service securely 
keeps track of the current state of \sjfs{} filesystems. 
It maintains a persistent map from unique filesystem identifiers $FS_{id}$ 
to their last two head commitments $\mathit{head}_{i_0,v-1}, \mathit{head}_{i_1,v}$ and 
their last tail commitment $\mathit{tail}_j$. 
It supports the following API. 
All requests are synchronous: they return only after completion.

\smallskip\noindent{\bf{(1) Session Establishment.}} 
A TEE mounting a filesystem first calls $\mathit{registerSession}$($\mathit{FS}_{id}, \mathit{TEE}_{id}, \mathit{credentials})$ to open a secure, mutually-authenticated channel to its designated \registry{} server. 
In addition to $\mathit{FS}_{id}$ (e.g. a long random bitstring constant across sessions), 
the client provides an identifier $\mathit{TEE}_{id}$ for the TEE instance mounting the disk,
authenticated with {\it credentials}.
In practice, $\mathit{TEE}_{id}$ can be derived from an attestation report over a fresh key pair and $\mathit{FS}_{id}$,
while {\it credentials} include a fresh proof of possession of the private key.
\registry{} uses these materials to authenticate the client and verify its attestation
against the authorization policy to mount an existing filesystem (if $\mathit{FS}_{id}$ is already registered) 
or to create a new filesystem (if $\mathit{FS}_{id}$ is not yet registered).
\registry{} accepts only one active session at a time for each filesystem instance. 
Hence, accepting a new session-establishment request from a client will terminate 
any session with another client. 
This supports recovery by a new TEE after a crash, and ensures only one TEE at a time 
can modify the registered~\jfs{} state, thereby preventing forking attacks.

\smallskip\noindent {\bf{(2) State Recovery}.} 
When remounting an existing filesystem, the TEE then calls
$\mathit{authorizeRecovery}(v, h_{i,v})$ where 
$v$ is the highest version number of its uberblock and $h_{i,v}$ is its authenticator. 
\registry{} compares them with the last two registered head commitments 
$\mathit{head}_{i_0,v}, \mathit{head}_{i_1,v+1}$.
If it matches the first, it returns $\mathit{head}_{i_0,v}, tail_j$. 
If it matches the second, it returns $\mathit{head}_{i_1,v+1}, tail_j$ and discards the first. 
(The head commitment in the response indicates the WAL head index and authenticator to use for recovery,
as they may not be included in the uberblock.) 
Otherwise, it returns an error message indicating that the uberblock is stale. 

\smallskip\noindent{\bf{(3) State Update.}} 
The TEE then calls 
$\textit{headUpdate}$($head_{i_2,v_2}$) and $\textit{tailUpdate}$($tail_j'$) when writing a new version of the uberblock and appending new WAL blocks, respectively.
\registry{} keeps the last two head commitments $head_{i_0,v}$ and $head_{i_1,v+1}$, 
and verifies that 
the head index is non-decreasing ($i_0 \leq i_1 \leq i_2$) and the version 
number is incremented ($v_2 = v + 1$) before accepting head updates. 
Similarly, it keeps the latest tail commitment $tail_j$ and verifies that the tail index is increasing ($j < j'$) before accepting tail updates.

\subsection{Security Analysis}

We argue that \sjfs{} ensures integrity and freshness properties as defined in \S\ref{sec:file_system_model} in a modular manner: building on crash-consistency,  
we show how the addition of cryptographic authentication and commitment management eliminates undetected data forgeries, rollback, and replay attacks, even when the storage backend is fully malicious.

\noindent{\bf Integrity and Freshness.} Although motivated by crash-safety, the baseline design of journaled filesystems 
implemented by \jfs{}
is a great match for \sjfs{}.
The systematic use of strong authenticators when storing and loading data excludes undetected forgeries. In our model, this means that data-forgery events that may break \jfs{} integrity or freshness are replaced by fail-safe data corruption event.  

The WAL is a direct implementation of the commit sequence in our model. Each block appended to its tail corresponds to the atomic extension of the commit sequence with the sequence of updates recorded in this block. \jfs{} waits for completion of the WAL block write before acknowledging any pending synchronization event, as expected from a POSIX filesystem, and \sjfs{} further waits for registration of the WAL tail commitment. 
Assuming the WAL replay logic of \jfs{} is correctly implemented, this ensures \sjfs{} will remount the filesystem in the state obtained by applying all updates recorded in WAL blocks $(\wb_k)_{i \leq k < j}$ since the creation of the filesystem. Any attempt to remount the filesystem from an outdated persisted state will be detected and replaced by a fail-safe data corruption error.

Besides these mechanisms, the \jfs{} logic responsible for concurrency control, fine-grained synchronization, and POSIX compliance is  entirely implemented in TEE-protected main memory, which ensures  integrity, confidentiality, and freshness within each session. 

While this goes beyond the scope of this work, this argument could be formalized as a simulation of \projectname{} session runs (with havoc events, possibly ending by a crash) by a matching \jfs{} session run, 
with identical in-memory states, live event sequence, and commit sequence, except for the replacement of every adversarial event \emph{Forge(inode, addr, data)} with a benign data corruption event \emph{Corrupt(inode, addr)}, both leading to matching errors in \jfs{} and \sjfs{}, respectively.
Such an argument would also confirm that, from the application viewpoint, shielding does not 
trigger any other error or new behavior.

\noindent{\bf Recoverability.} We also build on the existing (informal) crash-consistency argument of \jfs{} to ensure that,
in the absence of data corruption, \sjfs{} can recover from the same persisted state as \jfs{}.
\begin{enumerate}
    \item Tail updates follow a \emph{persist-then-register} discipline, ensuring that recovery based on a registered tail authenticator will always find the corresponding WAL blocks persisted on disk. Conversely, persisted unregistered WAL tail blocks will safely be ignored at recovery time.

    \item Head updates follow a \emph{register-then-persist} discipline, ensuring that \sjfs{} can always recover from the same highest uberblock version as \jfs{}, relying on the same persisted blocks in the storage pool. 
    This comes at the cost of keeping the last \emph{two} head commitments, so that we always persist version $v$ of the uberblock before registering version $v+1$. 
    In case of a crash between registration and persistence of version $v+1$, recovery 
    will safely proceed from version $v$, ignoring the registered head for version $v+1$. 
\end{enumerate}

\noindent{\bf Deadlock-Freedom and Liveness.}
\sjfs{} manages and persists a small amount of additional metadata for integrity and freshness, but it does not otherwise affect the operations of the filesystem.
It additionally registers the logical state of the whole filesystem at every synchronization point triggered by the application, using a protocol that may incur additional latency but is itself deadlock-free as long as \registry{} is live. 
Since this additional synchronization occurs either immediately before or immediately after 
a point where \jfs{} was already waiting
for the lower-layer completion on a block write, this does not introduce new deadlocks (although it may affect potential existing race conditions present in \jfs).

\noindent{\bf \registry{} fault model.} If \registry{} is unavailable (e.g., during disaster recovery),
synchronous updates and ownership transfers are blocked, i.e., a filesystem cannot be mounted
from another VM, including recovery after a crash.
Asynchronous operations within an already established session continue with no impact on performance or security.
Thus, \registry{} unavailability affects liveness, but not safety.

\noindent{\bf Authorization Policies.}
In our threat model, \sjfs{} guarantees that the current state of a given filesystem is the outcome of executing authorized application code in TEEs. While \registry{} enforces policies before authorizing a client to mount a filesystem, 
end-to-end security still critically depends on associating meaningful attestation-based policies
to filesystem instances, based for instance on a combination of code reviews, code signing, and transparency services. 

As a positive example, the application can be programmed in such a way that, after mounting its filesystem, it always logs its own attestation report and configuration in an append-only access-controlled file stored in the filesystem itself (and fsyncs its update) before starting normal operation. 
As long as this programming discipline is enforced by any application code that passes the attestation policy associated with a filesystem identifier, the integrity and completeness of the resulting audit log is guaranteed for the lifetime of the filesystem.

\noindent{\bf Confidentiality.} 
While \sjfs{} focuses on integrity and freshness, 
it can also achieve confidentiality by standard means
already available in TEEs and \jfs{}. 
The TEE protects any data 
in memory, and this protection can be extended by encrypting all blocks 
at rest (using, e.g., AES-XTS) with a symmetric key released to the TEE 
after verifying its attestation against a suitable policy.

%% file: implementation.tex
\section{Implementation} \label{sec:implementation}

\noindent{\bf \projectname{}.} We base our implementation on ZFS 2.3-release~\cite{openzfs_github}.
We add 2.5K lines of code, implementing the security refinements described in \S\ref{sec:architecture} 
without affecting its core invariants for POSIX compliance, consistency and, recoverability.
We do not need to modify the ZFS persistent data layout and I/O pipelines.
We adapt the ZIL logging flows, which rely on a single thread for ordering the blocks to be persisted at the ZIL tail. 
We extend the local state of the ZFS I/O handles that write these blocks 
(running on other threads) to incrementally compute and store their authenticators in their headers,
without the need for shared in-memory data structures.
For recovery, ZFS replays ZIL blocks sequentially, so we verify their authenticators incrementally as part of that flow.  

To register commitments from existing ZFS threads, we implement kernel-to-user IPC using netlink sockets~\cite{netlink} and auxiliary user-space threads that relay their messages to \registry{} over QUIC, as explained next.  

\smallskip\noindent{\bf \registry.} We implement the commitment registry by adapting  
the CCF distributed ledger~\cite{10.14778/3626292.3626304}. CCF provides a TEE-protected transactional key-value store in the confidential computing threat model. 
It achieves consistency and fault-tolerance using a consensus protocol between replicas hosted on multiple SEV-SNP nodes, where each node maintains its own copy of the KV store in memory
and its own copy of the ledger on disk (logging all transactions for auditing purposes). 

In contrast with prior works that either rely on single-instance fallible TEEs or TPMs~\cite{11023313}, or simply assume a reliable trusted third party, CCF meets our availability and fault-tolerance requirements
through tunable replication. As long as a majority of replicas are functioning, 
it tolerates up to $f$ independent failures at a time when configured to run on $2f+1$ nodes. 
(New nodes can be added to replace failed nodes without affecting safety or liveness.)
 Compared with classical Byzantine fault-tolerant protocols 
such as Practical Byzantine Fault Tolerance (PBFT)~\cite{castro1999practical}, which would require $3f+1$ nodes, 
CCF uses simpler, lower-latency protocols leveraging our assumption that each of its attested TEEs is fail-safe.

We adapt CCF to reduce latency in two ways: (1) since \projectname{} already provides stable storage, 
we keep commitments in the KV store, but we do not persist their history in the ledger. 
(2) We replace the network layer of CCF to use Linux sockets and HMAC-SHA-256 for 
authenticated communication between replicas, 
and we build a client-server library based on the ngtcp2~\cite{ngtcp2_github}
implementation of the QUIC protocol for authenticated sessions between \projectname{} and \registry{}. 

\registry{} can keep track of the current commitments for many instances of \projectname, 
since their \registry{} state is minimal: 216 B per filesystem, 
assuming 32 B SHA-256 authenticators and 8 B identifiers. 
Hence, a \registry{} instance whose nodes each have 1 GiB available for commitment state can hold commitments for at least 5 million \projectname{} clients. 

\smallskip\noindent{\bf Block-tampering framework.} To evaluate \projectname{}'s security,
we implement a framework capable of simulating adversarial edits to persistent storage
 to check if the filesystem detects the attack. Unlike other tools that introduce random
errors, the tool takes as user-provided input both the type of attack (e.g., rollback, replay, or equivocation) and
the target.

The framework is a user-space application on top of the filesystem-specific driver.
Recall that the in-memory state is protected by the TEE; therefore, only persistent state
is vulnerable to tampering.
Our framework operates on unmounted disks by rebuilding the in-memory state,
modifying the target WAL blocks in-memory, and writing
the altered blocks back to disk.
Our implementation focuses on ZFS, using {\tt libzfs} on the POSIX block device interface
of the target disk to find the ZIL head, parse the ZIL to locate the target block to edit,
and write back the tampered blocks.

As ZFS implements checksums for its data and metadata blocks,
arbitrary modifications to block payloads are likely to be detected. While prior work~\cite{Kadav2007ReliabilityAO} shows that ZFS can detect random corruption
in its Merkle tree-based storage pool and ZIL, our tool shows that ZFS is unable to
detect more advanced attacks that also correct block checksums. In contrast, the cryptographic hashes used for authenticators
in \projectname{} are effective at preventing such attacks. $\S$\ref{subsec:security} presents the supported types of attacks and resulting findings.

%% file: evaluation.tex
\section{Evaluation}
\label{sec:evaluation}

We evaluate \sjfs{} across three dimensions: 
performance (\S\ref{subsec:performance}, \S\ref{subsec:hybrid_design}), scalability (\S\ref{subsec:scalability}), and security (\S\ref{subsec:security}).

\subsection{Experimental Setup and Methodology}

\noindent \textbf{Hardware.}
We deploy \sjfs{} and \registry{} in AMD Gen3 CPU-based VMs (Azure Dadsv5) and Confidential VMs (CVMs) (DCadsv5).
\sjfs{} runs on VMs (or CVMs) with 32 vCPUs and 128 GB of RAM while \registry{} runs on VMs (or CVMs) with 8 vCPUs and 32 GB of RAM. 
The round-trip network latency between VMs is 0.1ms (p99) and between CVMs is 0.3ms (p90) and 0.6ms (p99). The higher
network latency in CVMs is due to the lack of support for accelerated networking. 
We expect cloud providers will soon support SR-IOV-based passthrough of confidential-computing-enabled NICs (such as Azure Boost~\cite{azure_boost} and NVIDIA Bluefield~\cite{bluefield} DPUs) to CVMs. 
We run the experiments on both VMs and CVMs to evaluate the performance of \sjfs{} with and without SEV-SNP.

VMs and CVMs hosting \sjfs{} are each assigned three locally redundant (LRS) Azure Premium SSD Managed Disks, each with 40 GB capacity, 7500 IOPs, and 250 MBps throughput, for a single filesystem with no replication.
This reflects a realistic high-performance cloud setup where the storage is remote and untrusted.

\noindent{\bf{Baseline Systems.}} We compare \sjfs{} against: 
(1) {\bf ext4}, detecting data corruption (bit rot, misdirected writes) using plain checksums (i.e., CRC-32); 
(2) {\bf ext4 with dm-integrity}~\cite{dm_integrity}, 
using a RAID-0 device and the dm-integrity device-mapper
to provide block-level authentication using cryptographic checksums (SHA-256); 
(3) {\bf ZFS}, detecting data corruption using checksums (i.e., Fletcher-4); 
and
(4) {\bf \sjfs{} without \registry{}}, providing session integrity and freshness given trusted initial commitments. 

\noindent{\bf{Baseline Applications.}} We evaluate their performance using fio~\cite{fio} 
as well as real-world applications: 
(1) a database workload: a PostgreSQL database over TPC-C workload~\cite{cloudsuite_data_serving}
with two configurations of 10 and 100 Warehouses; 
(2) filebench~\cite{filebench, filebench_github} with varmail and fileserver workloads; and 
(3) a key-value store workload: RocksDB~\cite{rocksdb_github}.

\noindent{\bf{Evaluation Highlights.}} Our evaluation shows that \sjfs{} provides integrity and freshness with performance similar to baseline ZFS across almost all real-world workloads. It is $1.7\times$ slower in the worst case, an fsync-intensive workload (e.g., filebench varmail), while it incurs negligible overheads ($<10\%$) for the rest (TPC-C, RocksDB).
Overall, ext4 achieves higher throughput than ZFS and \sjfs{} across most workloads, largely due to its simpler metadata structures and lower write amplification for asynchronous updates. 
In contrast, ZFS trades some performance for significantly stronger integrity and consistency guarantees, such as end-to-end checksumming and logging all updates through a write-ahead log. Importantly, for fsync-intensive workloads that emphasize durability, \sjfs{} outperforms ext4 and ext4 with dm-integrity by up to $1.6\times$.
\subsection{Performance}
\label{subsec:performance}

\begin{figure*}[htbp]
  \centering
  \includegraphics[width=0.85\textwidth]{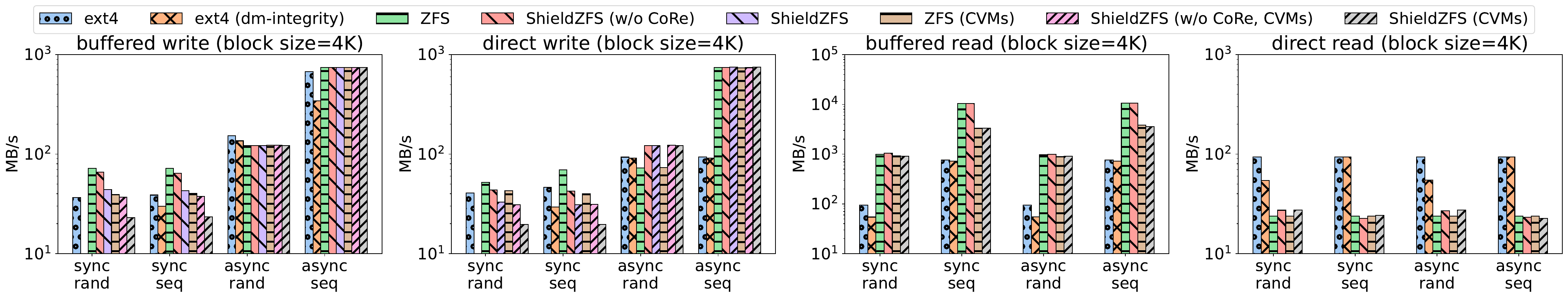}
  \vspace{-1.em}
  \includegraphics[width=0.85\textwidth]{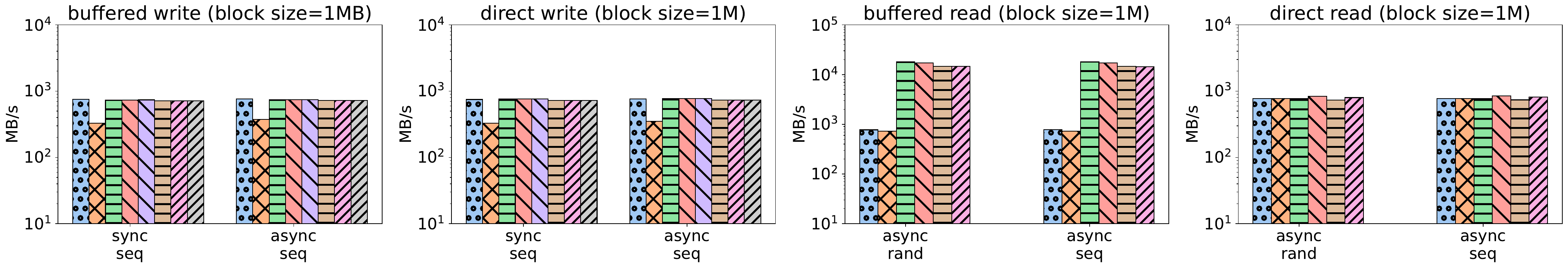}
  \caption{fio read and write throughput with 4 KB and 1 MB blocks.}
  \label{fig:write-read-throughput}
\end{figure*}

\noindent{\bf{Methodology.}} {\underline{fio}}: we evaluate \sjfs{} read and write performance 
in configurations with varying parameters: 
(i) read or write,
(ii) sequential or random, 
(iii) direct or buffered writes (in direct reads the page cache is disabled, isolating the disk access time),
(iv) durability: synchronous or asynchronous,
(v) block size: small (4 KB) or large (1 MB).
We set iodepth to 1 for synchronous operations and 32 for asynchronous operations.
We run each configuration for 30 seconds (ramp-up time) and then take measurements for 120 seconds.  

\noindent{{\underline{TPC-C}}}: we benchmark TPC-C using a dockerized version of PostgreSQL~\cite{cloudsuite_data_serving}. 
The client process runs on a separate VM.
The clients and the database server are not CPU-bound (10\% and up to 80\% usage respectively).
We run 60 clients for 10 Warehouses (10W) and 90 clients for 100 Warehouses (100W); 
these configurations yield database sizes around 1GB and 10GB, respectively. 
We measure the throughput (transactions per second) and average latency (ms) of transactions. 
The experiments comprise a loading phase, a 2-minute ramp-up phase, and a 6-minute measurement phase.  

\noindent{{\underline{Filebench}}}~\cite{filebench, filebench_github}: we benchmark varmail, 
a mail server workload that creates many small files and performs frequent fsyncs (every 4 syscalls);
it stresses metadata operations and synchronous writes, 
making it a good benchmark to evaluate the cost of fsync. 
We also benchmark fileserver with a mixed I/O pattern and moderate file sizes. 
We use their default configurations, and measure both throughput (IOPS) 
and average latency (ms) of their operations.

\noindent{{\underline{RocksDB}}}: we use the integrated benchmarking tool~\cite{rocksdb_benchmarks} 
with the default configuration. All workloads execute asynchronous writes and reads.
Periodically, RocksDB performs background compactions to merge and flush  
the SST files to persistent storage. RocksDB LSM size is 9 GB.
We measure the throughput (ops/s) for six workloads: \emph{(1)} bulkload: loads the database with KV pairs in random order, 
\emph{(2)} readrandom: randomly reads existing keys, 
\emph{(3)} overwrite: randomly overwrites keys into the database,
\emph{(4)} readwhilewriting: spawns one writer and multiple reader threads, 
\emph{(5)} fillseq\_wal\_enabled: sequentially fills the database with RocksDB-WAL, 
\emph{(6)} fillseq\_wal\_disabled: sequentially fills the database with no RocksDB-WAL.

\begin{figure*}[htbp]
  \centering
  \includegraphics[width=0.85\textwidth]{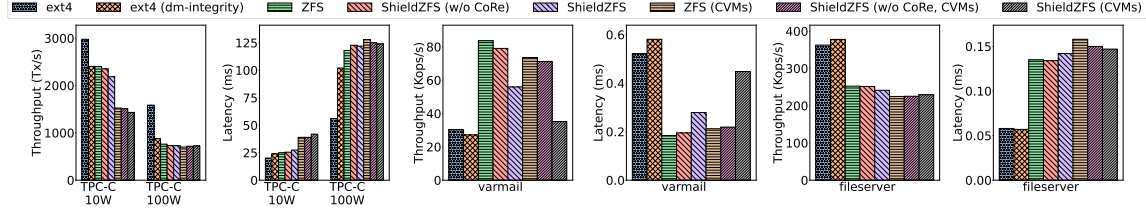}
  \vspace{-1.em}
  \caption{Throughput/latency graphs for TPC-C (left panels), varmail (middle panels), and fileserver (right panels).}
  \label{fig:applications-throughput}
\end{figure*}

\noindent{\bf{Results.}} {\underline{fio}} (Figure~\ref{fig:write-read-throughput}). 
We first discuss 4K blocks, which stress metadata management, then more briefly 1M blocks and the deployment in CVMs.

Compared to ZFS, \sjfs{} performs similarly for asynchronous writes, and incurs negligible overheads for reads. 
This is expected, since writes are buffered in ZFS cache and flushed to disk without blocking the application,
whereas reads rely on metadata cached in protected memory and amortize their authentication costs. 
On the other hand, \sjfs{} is slower up to $1.7\times$ and $2.3\times$ for buffered and direct synchronous writes respectively, 
highlighting the cost of producing and registering tail commitments.
(\sjfs{} without \registry{} is slower up to $1.09\times$ and $1.6\times$ for buffered and direct synchronous writes respectively compared to ZFS,
isolating the cost of tail commitment computations.) 
Registration is a bottleneck also for latency: fio reports average synchronous-write latencies 
of~$1 ms$ for ZFS, $1.13 ms$ for \sjfs{} without \registry{}, and $1.8 ms$ for \sjfs{},
which is consistent with average registration latencies of $0.7 ms$ for a single instance in \S\ref{subsec:scalability}.  

We also compare \sjfs{} to dm-integrity. With dm-integrity, ext4 has $1.5-5\times$ slower write throughput
and $1.7\times$ slower throughput for random reads, which
reflects the cost of 
cryptographic operations (SHA-256) and storing/fetching integrity metadata from disk.
(The overhead is much lower for sequential reads since they amortize integrity metadata reads.)  
On the one hand, \sjfs{} is $1.14-1.65\times$ faster than ext4 with dm-integrity for all synchronous write workloads
and even $1.17\times$ faster than plain ext4 synchronous buffered writes.
On the other hand, \sjfs{} is almost $4.5\times$ slower for direct reads, possibly reflecting ZFS more complex metadata structures. 
Buffered reads reflect filesystems' cache fill and readahead approach, with \sjfs{} performing similarly to ZFS and outperforming, in some cases, ext4 with dm-integrity by up to $10\times$.

With 1 MB blocks, the performance of all filesystems converges: they fully saturate disk bandwidth and device utilization (91--99\%) for both reads and writes.
fio reports average latencies of $42$--$44 ms$ for synchronous writes for all systems with the exception of 
ext4 with dm-integrity ($96 ms$).
This shows that registration costs become negligible. 
However, ext4 with dm-integrity
is still $2.3\times$ slower than plain ext4 for writes and saturates disk bandwidth only for reads. 

Execution of ZFS and \sjfs{} in CVMs incurs negligible overheads for the reads and asynchronous writes, except for the case of sequential reads
where CVMs-based execution is up to $3\times$ slower than VMs-based execution.
For synchronous writes, CVMs slow down the performance of \sjfs{} 
by up to $2-2.5\times$ w.r.t. VMs-based execution.

\begin{figure*}[htbp]
  \centering
  \includegraphics[width=0.85\textwidth]{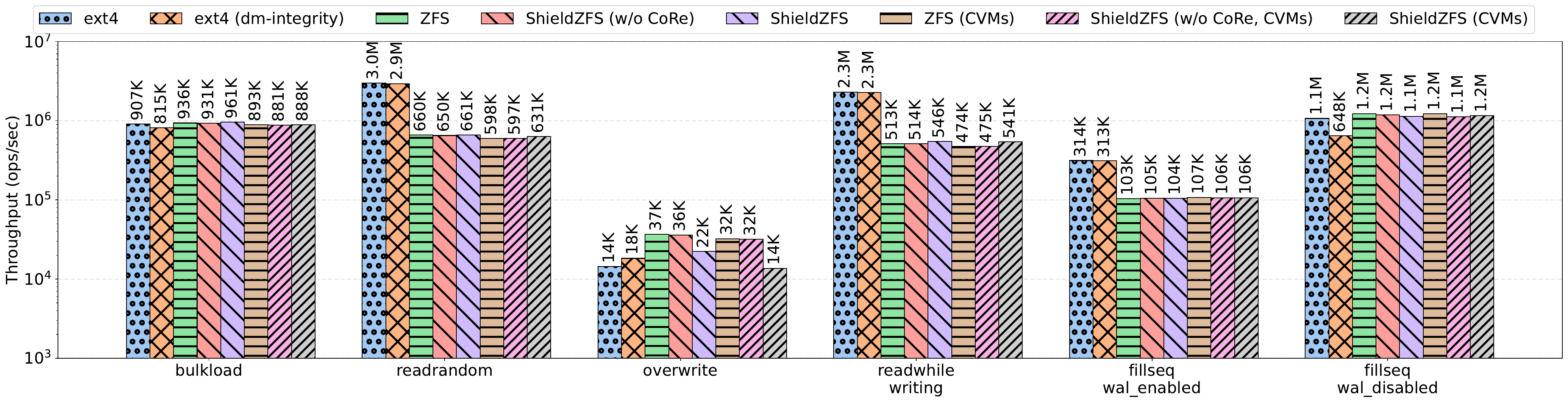}
  \vspace{-1.em}
  \caption{RocksDB throughput across various workloads.}
  \vspace{-1.em}
  \label{fig:rocksdb-throughput}
\end{figure*}

\noindent{\underline{TPC-C}} (Figure~\ref{fig:applications-throughput}, left).  Compared to ZFS, \sjfs{} has $1.05\times$ slower throughput
 for both 10W and 100W,
which we attribute to registration costs:
without \registry, \sjfs{} performs almost as ZFS, with a 0.3$\%$ throughput 
overhead for both workloads, suggesting the cryptographic overhead of commitments is negligible.

Compared to ext4, ZFS is $1.2\times$ and $2\times$ slower (in throughput) for 10W and 100W, respectively. This is expected since TPC-C is a write-heavy latency-sensitive workload 
with small transactions and frequent commits, making  
ext4 a better choice for pure performance. 
With dm-integrity, ext4 incurs $1.2\times$ and $1.9\times$ slower throughput (and 
higher latency) for 10W and 100W, respectively, much closer to baseline ZFS performance.

The impact of CVMs is negligible for TPC-C with 100W, possibly due to the larger database size and more frequent I/O, whereas
for 10W, \sjfs{} is $1.5\times$ slower than VMs-based execution, due to the higher commitment registration latency.

\noindent{\underline{Filebench}} (Figure~\ref{fig:applications-throughput}, middle and right). For varmail, \sjfs{} is slower than ZFS by $1.7\times$; this overhead is 
attributed to the fsync-intensive nature of varmail.
Without \registry{}, this overhead drops to 2.5\%,  
confirming registration is the main bottleneck.  \sjfs{} still
outperforms ext4 and ext4 with dm-integrity by $1.6\times$. Execution in CVMs incurs up to $2\times$ overheads in \sjfs{}
due to the higher commitment registration latency.

For fileserver, a read-heavy workload with large sequential asynchronous reads/writes, \sjfs{} performs similar to ZFS, while ext4 outperforms both by $1.5\times$.
CVMs have negligible impact on performance for this workload.

\noindent{\underline{RocksDB}} (Figure~\ref{fig:rocksdb-throughput}). 
\sjfs{} performs as well as ZFS on most workloads, 
which mostly perform bulk sequential I/Os and are thus unaffected by commitment costs,
whereas it is $1.68\times$ slower on the overwrite workload, 
which exercises random persisted key writes. All variants of ZFS are consistently slower than ext4 (with faster asynchronous operations). ext4 with 
dm-integrity introduces modest overheads, $\approx13\%$ over ext4. 
ext4 with dm-integrity performs worse in the fillseq\_wal\_disabled workload, where entire tables (64MB) are compacted and written sequentially, 
but performs equally well in workloads that are read-dominant, such as readwhilewriting and readrandom.
CVMs have negligible impact on performance.

\smallskip\noindent{\bf{Recovery.}}
We measure the impact of \sjfs{} on the time to remount a filesystem. 
After a clean unmount, the WAL is empty, and the additional costs for \sjfs{} are negligible (a roundtrip to \registry{} and a single hash computation).
After a crash, however, many WAL blocks may need to be replayed, along with their hash computations and verifications. We evaluated \sjfs{} 
for a WAL with 500 and 2500 blocks. \sjfs{} incurs 1.3$\times$ overhead compared to original 
ZFS (0.66s compared to 0.52s for 500 blocks and
1.80s compared to 1.41s for 2500 blocks).

\subsection{Hybrid Design for Integrity and Performance}
\label{subsec:hybrid_design}
Some modern data stores~\cite{227798, 277820, 8418608} are designed to
provide integrity at the application level.
For instance, variants of RocksDB~\cite{227798, 277820} maintain
a succinct Log of authenticators for all their SSTables (DB directory),
and this Log suffices to provide crash-consistency and durability guarantees after  (synchronous) Put operations.
For best performance, we can leverage these mechanisms, use \sjfs{} specifically to protect the Log, and use, e.g,. ext4 to store the rest of the DB.  We use RocksDB to show this approach empirically below.

\smallskip\noindent{\bf{Methodology.}} We mount RocksDB's DB directory on ext4 to optimize for asynchronous Get and Put operations,
and the Log on \sjfs{} for durability and freshness.

\begin{figure}[t]\scriptsize
\centering
\begin{tabular}{@{}llrr@{}}
\toprule
\textbf{DB dir} & \textbf{Log} &  \textbf{Get/Put} &  \textbf{Durable Put} \\
\midrule
ext4       & ext4       & 1.788\,M ops/s & 0.442\,K ops/s \\
zfs        & zfs        & 0.368\,M ops/s & 1.093\,K ops/s \\
ShieldZFS  & ShieldZFS  & 0.367\,M ops/s & 0.724\,K ops/s \\
ext4       & zfs        & 1.727\,M ops/s & 1.087\,K ops/s \\
ext4       & ShieldZFS  & 1.715\,M ops/s & 0.715\,K ops/s \\
\bottomrule
\end{tabular}
\caption{Throughput for different configurations of RocksDB.}
\label{fig:hybrid_system_throughput}
\vspace{-1ex}
\end{figure}

\smallskip\noindent{\bf{Results}} (Figure~\ref{fig:hybrid_system_throughput}). 
The all-ext4 configuration achieves the highest Get/Put throughput (1.788M ops/s) but suffers from poor Durable Put performance (0.442K ops/s).
Conversely, all-ZFS and all-\sjfs{} provide strong integrity guarantees and achieve much higher Durable Put throughput (1.093K and 0.724K ops/s),
but sacrifice asynchronous performance (0.368M and 0.367M ops/s). The hybrid approach (ext4 DB + \sjfs{} Log) is the optimal.

\subsection{Scalability}
\label{subsec:scalability}
We evaluate the scalability of \registry{}, i.e., its ability to support concurrent active filesystems. 
(Recall that \registry{} can hold commitments for many inactive filesystems.)

\begin{figure}[htbp]
  \centering
  \includegraphics[width=0.3\textwidth]{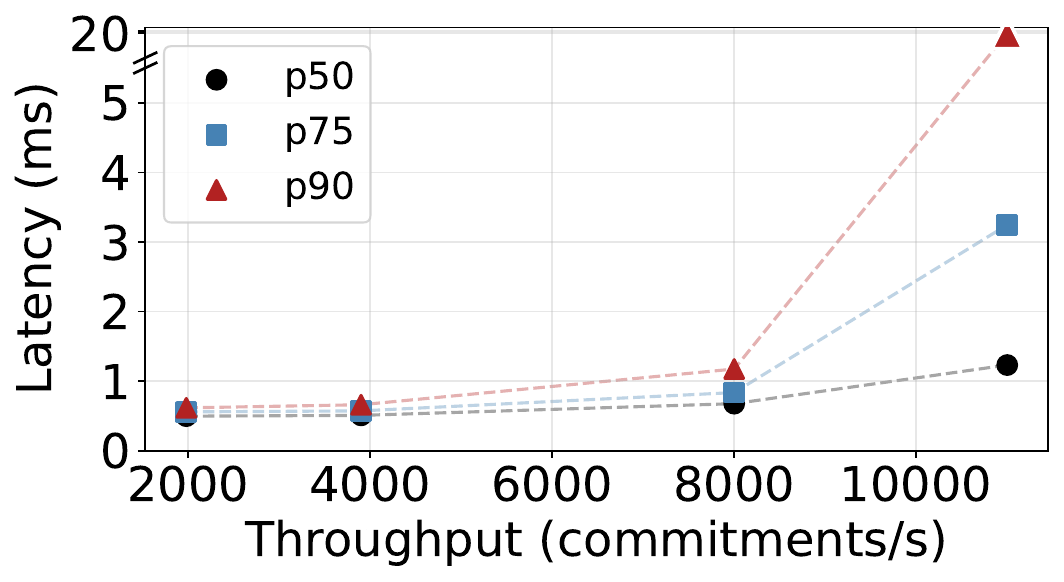}
  \vspace{-1.em}
  \caption{\registry{} latency and throughput.}
  \label{fig:core_scalability}
    \vspace{-1.em}
\end{figure}

\smallskip\noindent{\bf{Methodology.}} We spawn a varying number of user-space clients that register commitments continuously for 3 minutes stress-testing the \registry{}.

\smallskip\noindent{\bf{Results}} (Figure~\ref{fig:core_scalability}).
\registry{} p90 registration latency is $\leq 0.7 ms$ for up to 5K commitments/s, supporting 25--500 concurrent \sjfs{} instances under RocksDB workloads (which generate 10--200 commitments/s).

We also measured latency for the crash-safe replication protocol within \registry{} 
which varies from $0.25 ms$ to $0.5 ms$ for regular VMs and from $0.5 ms$ to $0.9 ms$ for CVMs.

\subsection{Security}
\label{subsec:security}
\noindent{\bf{Methodology.}} We consider two classes of attacks, 
single-block attacks that tamper with a single ZIL block, 
and multi-block attacks that tamper with multiple ZIL blocks. 
Single-block attacks include the following variations: (i) payload-modification attacks, 
(ii) block-header modification attacks (e.g., tampering with locators, identifiers, and checksums or authenticators), (iii) append-new-block attacks, and (iv) empty-block attacks. 
Multi-block attacks include: (i) reordering existing blocks and (ii) removing existing blocks. 

With our block-tampering framework we combine these attacks and parameters to synthesize more complex attack scenarios. For example,   
we performed forking attacks in the \sjfs{}, that is, when an attacker creates two different versions of the ZIL trying to make a system
accept as valid either of the two. We executed a combination of these attacks and variations 
resulting in a total of 63 attack scenarios for \sjfs{} and 42 for ZFS.

\smallskip\noindent\textbf{Results.} ZFS fails to detect corruptions when block checksums are tampered together with data so that the new checksum corresponds to the updated identifier fields, including scenarios in which ZIL blocks are reordered, dropped, or appended. In contrast, \sjfs{} detects all such attacks. 
Interestingly, our tool uncovered a subtle flaw in an early prototype: 
while \sjfs{} correctly authenticated the ZIL using $\mathit{tail}_{j}$,
it failed to detect an attack where the contents of the first block (head of the ZIL) were rewritten, 
due to our (flawed) assumption that accessing this block from the uberblock would suffice to 
authenticate its contents.

%% file: related_work.tex
\section{Related Work}
Prior work has explored storage integrity and freshness protection in the confidential computing threat model.

\noindent{\bf Integrity-preserving block storage.} Several systems reduce the cost of integrity checking by deferring or batching verification. PAC~\cite{11023313} performs integrity checks asynchronously in the background; sNVMe-oF~\cite{chrapek2025snvmeofsecureefficientdisaggregated} applies Merkle tree updates asynchronously in smart, trusted storage devices; and DMT~\cite{10.5555/3724648.3724672} amortizes verification across distributed nodes. Compared to \projectname{}, these systems offer weaker guarantees since reads can (temporarily) see corrupt or stale data.

Rollbaccine~\cite{10.1145/3786693} is a device-mapper that provides integrity by intercepting all block I/Os and replicating metadata 
across a set of replica TEEs on synchronous writes. Replicas rely on TEE memory to protect metadata, and on high-throughput disks for recovery. Compared to \projectname{}, Rollbaccine is operationally more complex, as each VM needs to be coupled with additional replica VMs, 
and requires significantly higher compute/storage resources.
  
DiskShield~\cite{10.1145/3320269.3384717} provides integrity and freshness for disk storage by interposing block I/O in trusted SSD firmware, maintaining cryptographic metadata (e.g., Merkle tree) to detect block-level tampering, rollback, and replay attacks. This approach is tightly coupled with the storage hardware layer. In contrast, \projectname{}  operates at the filesystem layer and does not require any changes to storage hardware.

\noindent{\bf Integrity-preserving filesystems.} Intel Protected File System (IPFS)~\cite{ipfs} uses per-file Merkle trees for confidentiality and integrity. While IPFS detects tampering of file contents, it lacks a global authenticated structure across files and it is prone to rollback attacks when attackers replay older, valid versions of files together with their authentication metadata.

Nexus~\cite{8809505} ensures file-level confidentiality and integrity with high overheads ($2\times$ for many operations), while remaining susceptible to freshness and forking attacks. SecureFS~\cite{10.1145/3471621.3471840} is a bespoke filesystem that provides specific protection against freshness attacks by maintaining versions at the block (slab) level and propagating versioning through pointers. On the contrary, \projectname{} enhances state-of-the-art POSIX-compliant filesystems to be resilient against all integrity and freshness attacks ($\S$\ref{subsec:security}). 

SUNDR~\cite{10.5555/1251254.1251263}, Plutus~\cite{10.5555/1090694.1090698}, jVPFS~\cite{10.5555/2002181.2002213}, SiRiUS~\cite{sirius}, SNAD~\cite{270756}, Maat~\cite{10.1145/1362622.1362644}, and
PCFS~\cite{10.1007/978-3-642-29963-6-5} provide secure storage in terms of confidentiality, integrity, and authorization. However, they leave rollback attacks out of scope. StrongBox~\cite{Dickens2018StrongBoxCI} targets rollback protection but relies on an expensive in-memory Merkle tree and hardware persistent secure counters, which need to be updated on every write.

\noindent{\bf Integrity protection at the application-level.} CCF~\cite{10.14778/3626292.3626304} builds a key-value store and a distributed ledger in the confidential computing threat model. We adapt its consensus and replication protocol to implement \registry{}.

Nimble~\cite{angel2023nimble} offers a lightweight ledger service managed by 
TEE replicas for availability. It incurs an average update latency of $2.5 ms$,
while requiring significant effort to identify the state that needs protection and failure recovery.

SPEICHER~\cite{227798}, EnclaveDB~\cite{8418608}, Treaty~\cite{9833660}, and TWEEZER~\cite{277820} are database and key-value stores that use
authenticated persistent data structures and append-only logs for storage-level integrity and freshness, similarly to \projectname{}'s WAL. In contrast with workloads running on top of \projectname{}, these systems require substantial application-level modifications.